\documentclass{aa}
\usepackage{graphicx}
\usepackage{amsmath}
\usepackage{txfonts}
\usepackage{natbib}
\bibpunct{(}{)}{;}{a}{}{,}

\begin{document}

   \title{First analysis of solar structures in 1.21 mm full-disc ALMA image of the Sun}

   \author{R. Braj\v{s}a \inst{1}
           \and
            D. Sudar
           \inst{1}
           \and
           A. O. Benz \inst{2}
           \and
           I. Skoki\'{c} \inst{1}
           \and
           M. B\'arta 
          \inst{3}     
          \and          
          B. De Pontieu
    	  \inst{4,10}
	  \and
	  S. Kim
	\inst{5}
	\and
	A. Kobelski
	\inst{6}
	\and
	M. Kuhar \inst{2,7}
           \and
	M. Shimojo 
	 \inst{8,9}     
        	\and
	 S. Wedemeyer\inst{10}     
          \and
	S. White
          \inst{11}   
        	\and
	P. Yagoubov
	\inst{12}
	\and
	Y. Yan
	\inst{13}
                    }

   \institute{Hvar Observatory, Faculty of Geodesy,
              Ka\v{c}i\'{c}eva 26, University of Zagreb, 10000 Zagreb, Croatia
              \and
              University of Applied Sciences and Arts Northwestern Switzerland, Bahnhofstrasse 6, 5210 Windisch, Switzerland
              \and
              Astronomical Institute of the Czech Academy of Sciences, Fri\v{c}ova 298, 25165 Ond\v{r}ejov, Czech Republic
              \and
           Lockheed Martin Solar \& Astrophysics Laboratory, 3251 Hanover Street, Org. A021S, B. 252, Palo Alto, CA 94304, USA
          \and
             Korea Astronomy and Space Science Institute, Daejeon, Republic of Korea            
             \and
             Center for Space Plasma and Aeronomic Research (CSPAR), University of Alabama in Huntsville, Huntsville, AL 35805, USA
             \and
             ETH: Institute for Particle Physics, ETH Z\"urich, 8093 Z\"urich, Switzerland
              \and
             National Astronomical Observatory of Japan, 2-21-1 Osawa, Mitaka, Tokyo 181-8588, Japan
             \and
            Department of Astronomical Science, The Graduate University for Advanced Studies (SOKENDAI), 2-21-1 Osawa, 
            Mitaka, Tokyo, 181-8588, Japan 
              \and
              Institute of Theoretical Astrophysics, University of Oslo, Postboks 1029, Blindern, 0315 Oslo, Norway
             \and
           Space Vehicles Directorate, AFRL, 3550 Aberdeen Avenue SE, Bldg 427, Kirtland AFB,
		NM 87117-5776, USA
	   \and 
             European Southern Observatory (ESO), Karl-Schwarzschild-Strasse 2, 85748 Garching bei M\"unchen, Germany		
	   \and 
              NAO, Chinese Academy of Sciences, Beijing, China                   
                      }

\offprints{R. Braj\v{s}a, \email romanb@geof.hr}

   \date{Release \today}

\abstract 
{Various solar features can be seen in emission or absorption on maps of the Sun in the millimeter and submillimeter wavelength range. The recently installed Atacama Large Millimeter/submillimeter Array (ALMA) is capable of observing the Sun in that wavelength range with 
an unprecedented spatial, temporal and spectral resolution. 
To interpret solar observations with ALMA the first important step is to compare 
solar ALMA maps with simultaneous images of the Sun recorded in other spectral ranges. 
}
{The first aim of the present work is to identify different structures in the solar atmosphere seen in the 
optical, infrared and EUV parts of the spectrum 
(quiet Sun, active regions, prominences on the disc,  magnetic inversion lines,  coronal holes and 
coronal bright points) in a full disc solar ALMA image. 
The second aim is to measure the intensities (brightness temperatures) of those structures 
and to compare them with the corresponding quiet Sun level.  
}
{A full disc solar image at 1.21 mm obtained on December 18, 2015 during a CSV-EOC campaign 
with ALMA is calibrated and compared with full disc solar images from the same day in H$\alpha$ line, 
in He I 1083 nm line core, and with various SDO images 
(AIA at 170 nm, 30.4 nm, 21.1 nm, 19.3 nm, and 17.1 nm and HMI magnetogram). 
The brightness temperatures of various structures are determined by averaging over corresponding regions of interest 
in the calibrated ALMA image.  
 }
{Positions of the quiet Sun, active regions, prominences on the disc, magnetic inversion lines, coronal holes 
and coronal bright points are identified in the ALMA  image. 
At the wavelength of 1.21 mm active regions appear as bright areas (but sunspots are dark),  
while prominences on the 
disc and coronal holes are not discernible from the quiet Sun background, although having slightly less 
intensity than surrounding quiet Sun regions. Magnetic inversion lines appear as large, elongated dark 
structures and coronal bright points correspond to ALMA bright points. 
}
{These observational results are in general agreement  with sparse earlier measurements at similar 
wavelengths. The identification of coronal bright points represents the most important new result. 
By comparing ALMA and other maps, it was found that the ALMA image was oriented  properly and that 
the procedure of overlaying  the 
ALMA image with other images is accurate at the 
5 arc sec level. 
The potential of ALMA for physics of the solar chromosphere is emphasized. 
}

\maketitle

\section{Introduction}

The Atacama Large Millimeter/submillimeter Array 
(ALMA)\footnote{http://www.almaobservatory.org}
\footnote{https://www.eso.org/sci/facilities/alma.html}
is currently the world 
largest ground based astronomical facility, capable of observing almost all types of 
celestial objects including the Sun.
The main advantage of solar observations with ALMA is mapping of the solar 
chromosphere with an unprecedented spatial, temporal and spectral resolution
in the  wavelength range between 0.3 mm and 8.6 mm.
Solar measurements are currently limited to two observing bands at 
1.3 mm and 3 mm. 

An important characteristic of ALMA is its capability to be used as an approximately 
linear thermometer of the gas in the solar atmosphere \citep{Wedemeyer2016}. So, the measured brightness 
temperature is directly proportional to the gas temperature of the observed structure or 
layer in the solar atmosphere. Therefore, it is very important to develop numerical 
simulations of various solar atmosphere types as \it{a priori} \rm and \it{a posteriori} 
\rm supporting tools for solar observations with ALMA \citep{Wedemeyer2015} and to compare 
results with available measurements.
It was clear from the initial phase that ALMA can be pointed at the Sun and that it will become an important 
tool if it could be made available for solar physics. Various applications were described by 
\citet{Bastian2002},  \citet{Wedemeyer2007}, 
\citet{Loukitcheva2008},  \citet{Karlicky2011}, 
and \citet{Loukitcheva2015}.   
An extensive review of solar scientific topics to be addressed with ALMA, including 
quiet Sun, active regions and prominences, was prepared by \citet{Wedemeyer2016}.

We briefly describe some of the earlier solar measurements in the ALMA wavelength 
range   
(from $\lambda = 0.3$ mm to $\lambda = 10$ mm).  
Brightness temperature measurements of the quiet Sun were 
reviewed and compared with results of theoretical models by \citet{Gary1996}, 
by \citet{Loukitcheva2004, Loukitcheva2015}, and by \citet{Benz2009}. 
Further, the quiet Sun emission  
in the wavelength range from $\lambda = 0.85$ mm to $\lambda = 8$ mm
was measured by \citet{Bastian1993a},  \citet{White2006},  
\citet{Brajsa2007a, Brajsa2007b}, and  \citet{Iwai2017}.
In the wavelength range 0.7 mm to 5 mm various measurements give 
the quiet Sun brightness temperature in the range from 5000 K to 8000 K, 
as summarized by \citet{White2017}.  In spite of scattered results and measurement 
uncertainties, there is an average trend of the brightness temperature increase with wavelength, as expected.

Solar active regions were observed 
in the wavelength range from $\lambda = 0.35$ mm to $\lambda = 9.5$ mm
by \citet{Kundu1972},  
\citet{Pohjolainen1991}, 
 \citet{Bastian1993a}, 
\citet{Lindsey1995}, 
\citet{Silva2005},
\citet{White2006}, 
\citet{Iwai2015},
\citet{Iwai2016}, and    
\citet{Kallunki2017}.   
As expected, active regions at mm and sub-mm wavelengths generally appear 
as bright areas, although sunspot umbrae, if resolved, have lower brightness 
temperatures than the quiet Sun level at $\lambda \le 3.5$ mm \citep{Lindsey1995, 
Loukitcheva2014, Iwai2015} or have almost equal brightness temperature as the quiet Sun 
at $\lambda = 8.8$ mm \citep{Iwai2016}.

Solar prominences, observed as filaments on the disc, are generally seen as 
structures of lower brightness temperature than the quiet Sun at mm 
wavelengths. Such results were obtained by measurements 
in the wavelength range from $\lambda = 4$ mm to $\lambda = 8$ mm  
\citep{Kundu1978, Schmahl1981, Vrsnak1992}.    
According to the list of radio observations of filaments compiled and corrected by 
\citet{Raoult1979}, the intensity contrast (lower brightness temperatures of filaments)
with respect to the quiet Sun decreases with frequency, being $\approx$5\% at 3.5 mm. 
\citet{Hiei1986} performed measurements of filaments on the disc at $\lambda = 8.3$ mm and 
at $\lambda = 3.1$ mm and further improved results of \citet{Raoult1979} taking into 
account observational beam/convolution effects.  
In this way larger brightness temperature differences between filaments and quiet Sun 
were measured.
At higher frequencies, \citet{Bastian1993b} found negligible contrast (but still a lower 
brightness temperature than the quiet Sun level) at and
near the location of H$\alpha$ filaments at $\lambda = 0.85$ mm. 

At mm wavelengths coronal holes are mostly observed as regions having 
brightness temperatures slightly below the quiet Sun level, but local enhancements 
of radiation within their borders can also be seen. 
Measurements which support this general result were performed 
in the wavelength range from $\lambda = 3.0$ mm to $\lambda = 8.8$ mm 
by
\citet{Kundu1976}, 
\citet{Kosugi1986}, 
\citet{Pohjolainen1997}, 
\citet{Gopalswamy1999}, 
\citet{Pohjolainen2000b}, 
\citet{White2006}, and 
\citet{Brajsa2007a, Brajsa2007b}.
Moreover, local enhancements of radio emission in polar regions were measured by 
\citet{Urpo1987}, 
\citet{Riehokainen1998}, and 
\citet{Pohjolainen2000a} in the wavelength range 
from $\lambda = 3.5$ mm to 
$\lambda = 8$ mm.   

As we have seen, there is a relatively large number of previous solar studies 
at  mm and to a lesser extent at sub-mm wavelengths. 
However, the main problem/deficiency of these works are too low spatial resolution 
and poor calibration. With ALMA it is now possible to substantially improve observations 
of the Sun at mm and sub-mm wavelengths. 

Thermal bremsstrahlung can be considered as the dominant radiation mechanism 
of the quiet solar radio emission at mm and sub-mm wavelengths 
\citep[e.g.][]{Tapping1994, Bastian1995, Benz1997}. 
A strong magnetic 
field would be necessary  (in the range 3300 Gauss $-$ 6600 Gauss, for $\lambda = 8$ mm and 
28000 Gauss and more for $\lambda = 1.21$ mm), 
for thermal gyromagnetic (cyclotron)  radiation 
\citep[e.g.][]{Brajsa2009}. 
Such a large magnetic field needed for the short-mm wavelength 
gyromagnetic emission is highly improbable. 

Calculations of the brightness temperatures of the quiet Sun and chromospheric network 
\citep[e.g.][]{Chiuderi1983, Loukitcheva2004, Loukitcheva2006}
are  usually based on the VAL \citep{Vernazza1981}, FAL \citep{Fontenla1993}, 
and CS \citep{Carlsson1992, Carlsson1995, Carlsson1997, Carlsson2002}
models. 
Modeling of filaments radio emission (or absorption) is not straightforward, since 
the results are strongly frequency dependent, especially in the mm and sub-mm 
wavelength range 
\citep{Chiuderi1990, Chiuderi1991, Chiuderi1992, Engvold1994, Tandberg1995}.
Moreover, many subtle effects should be taken into account, e.g., the influence of the 
prominence-corona transition region, or the visibility of filament channels and 
coronal condensations. 
Coronal condensations are structures lying over  magnetic inversion  lines which could 
be discernible at mm wavelengths, but remain invisible in the H$\alpha$ filtergrams 
\citep{Kundu1978, Vrsnak1992}.

Our analysis is divided in two parts: the observational part (present work, Paper I) and the 
modeling part (subsequent paper, Paper II). In present paper we first describe measurements 
and methods 
of data reduction (Sect. 2), then present the results of the observational analysis of the 
ALMA image and compare them with other data (Sect. 3) and finally discuss the most 
important results and summarize conclusions (Sect. 4).

\section{Measurements and data reduction}

\begin{figure*}
\centerline{\includegraphics[width=0.45\textwidth]{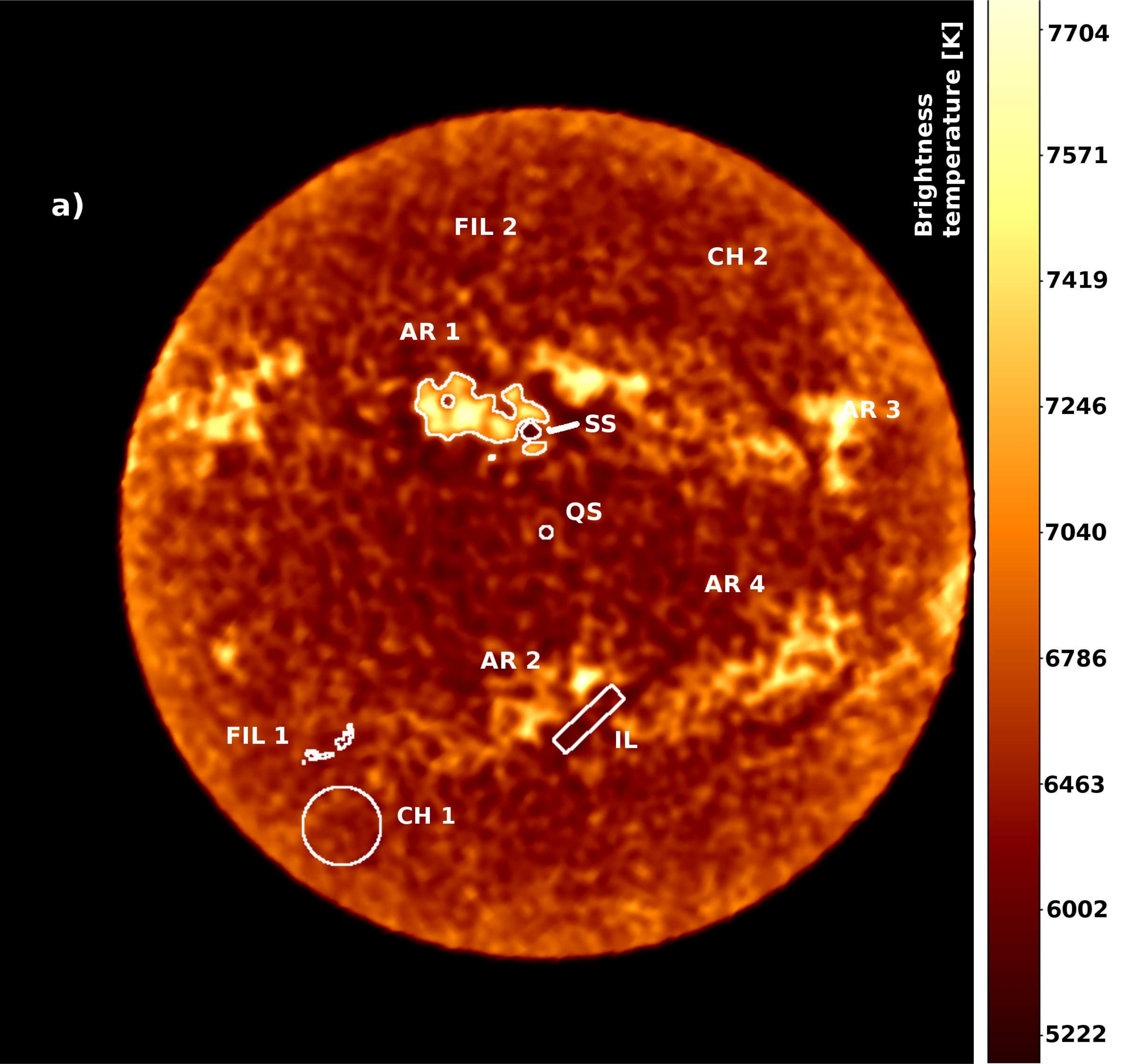}\includegraphics[width=0.45\textwidth]{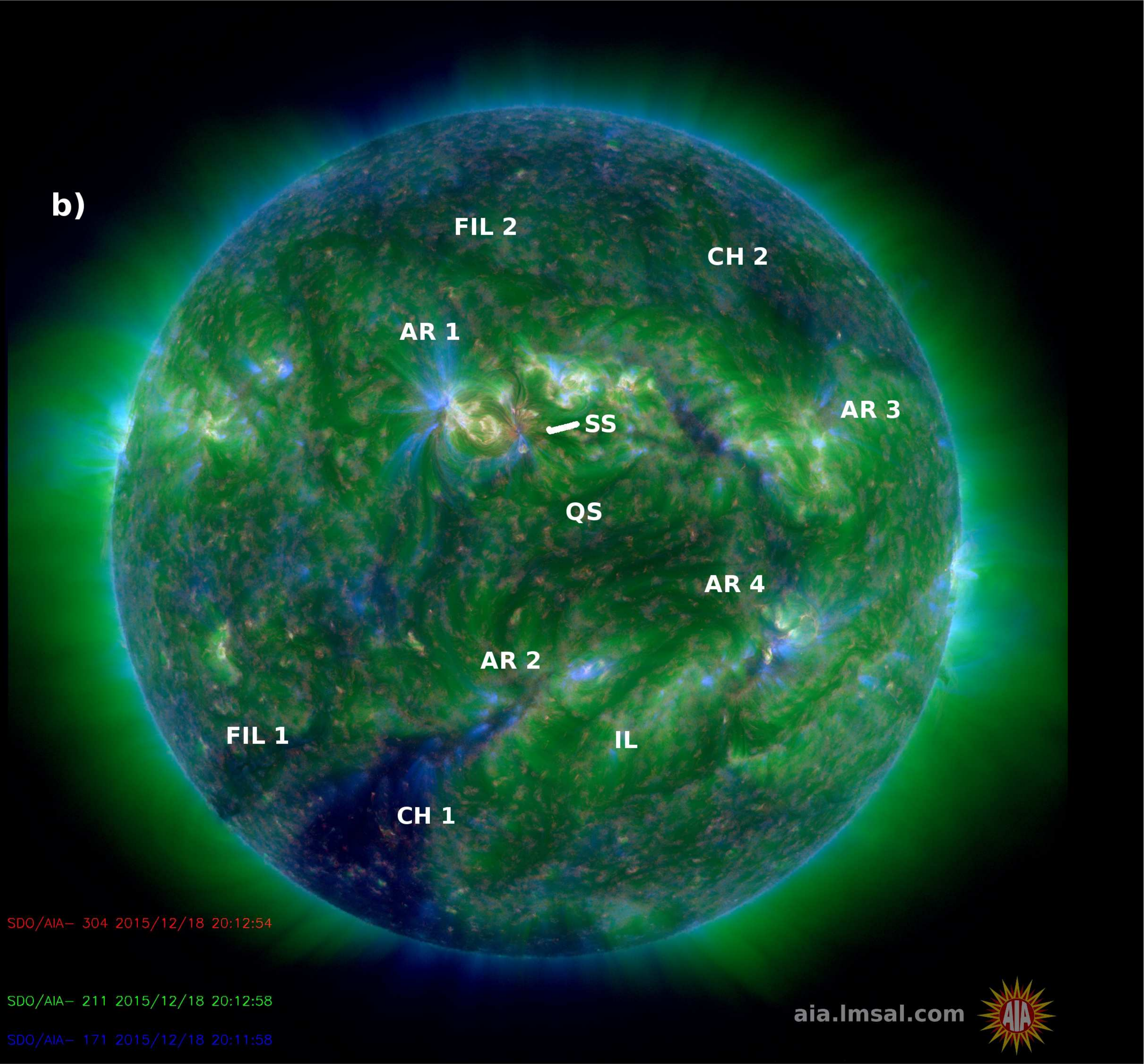}}
\centerline{\includegraphics[width=0.45\textwidth]{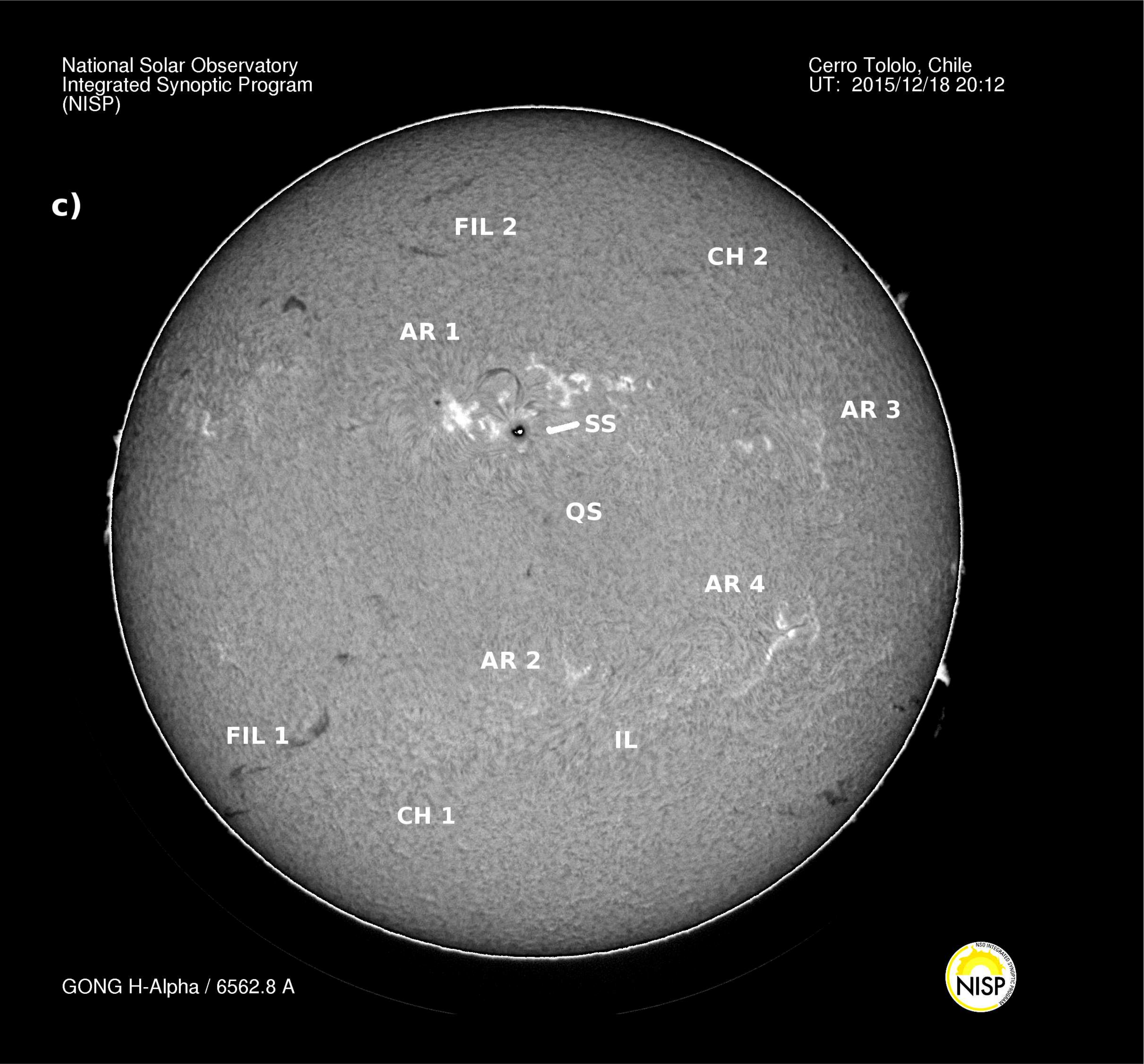}\includegraphics[width=0.45\textwidth]{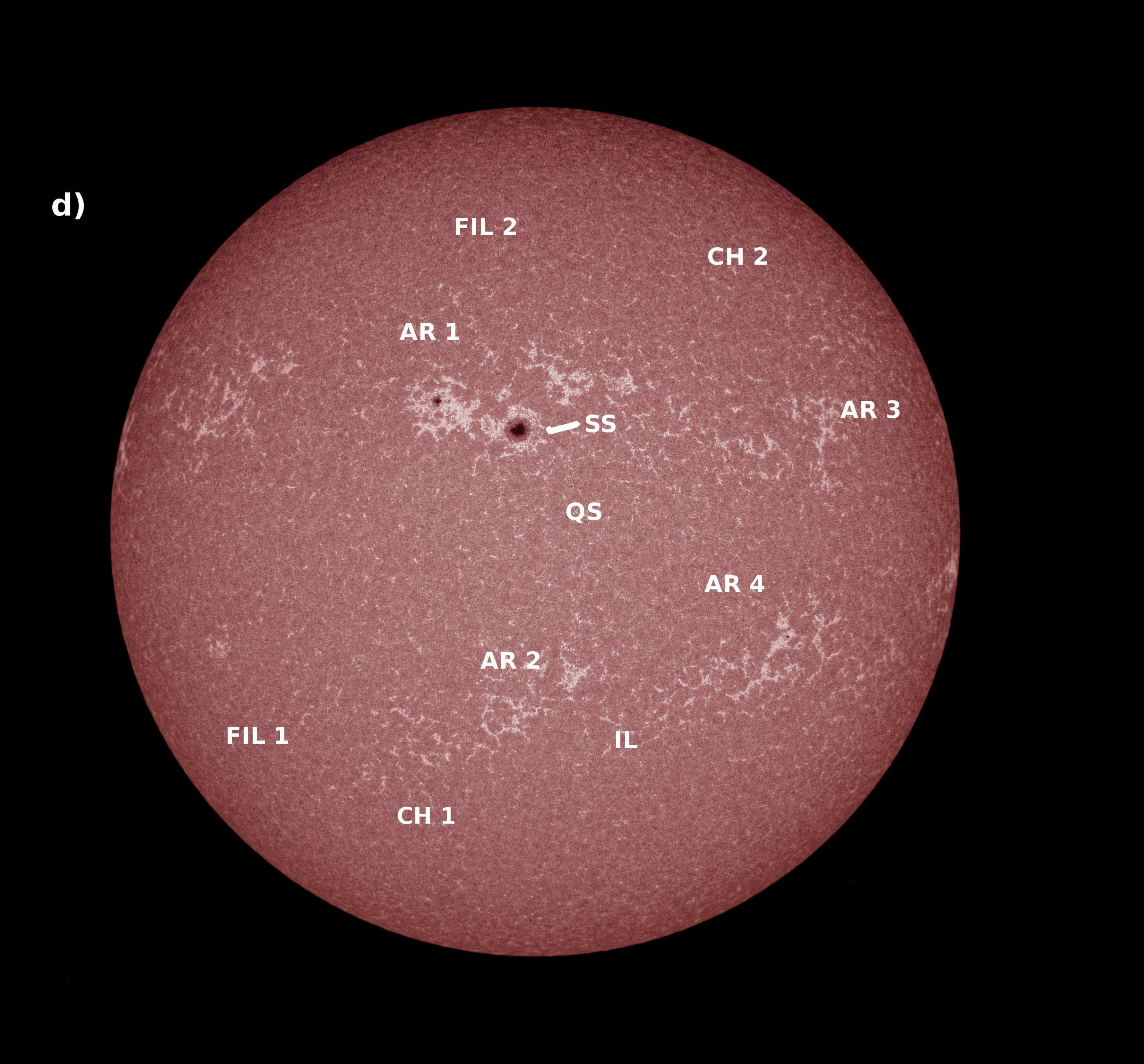}}
\centerline{\includegraphics[width=0.45\textwidth]{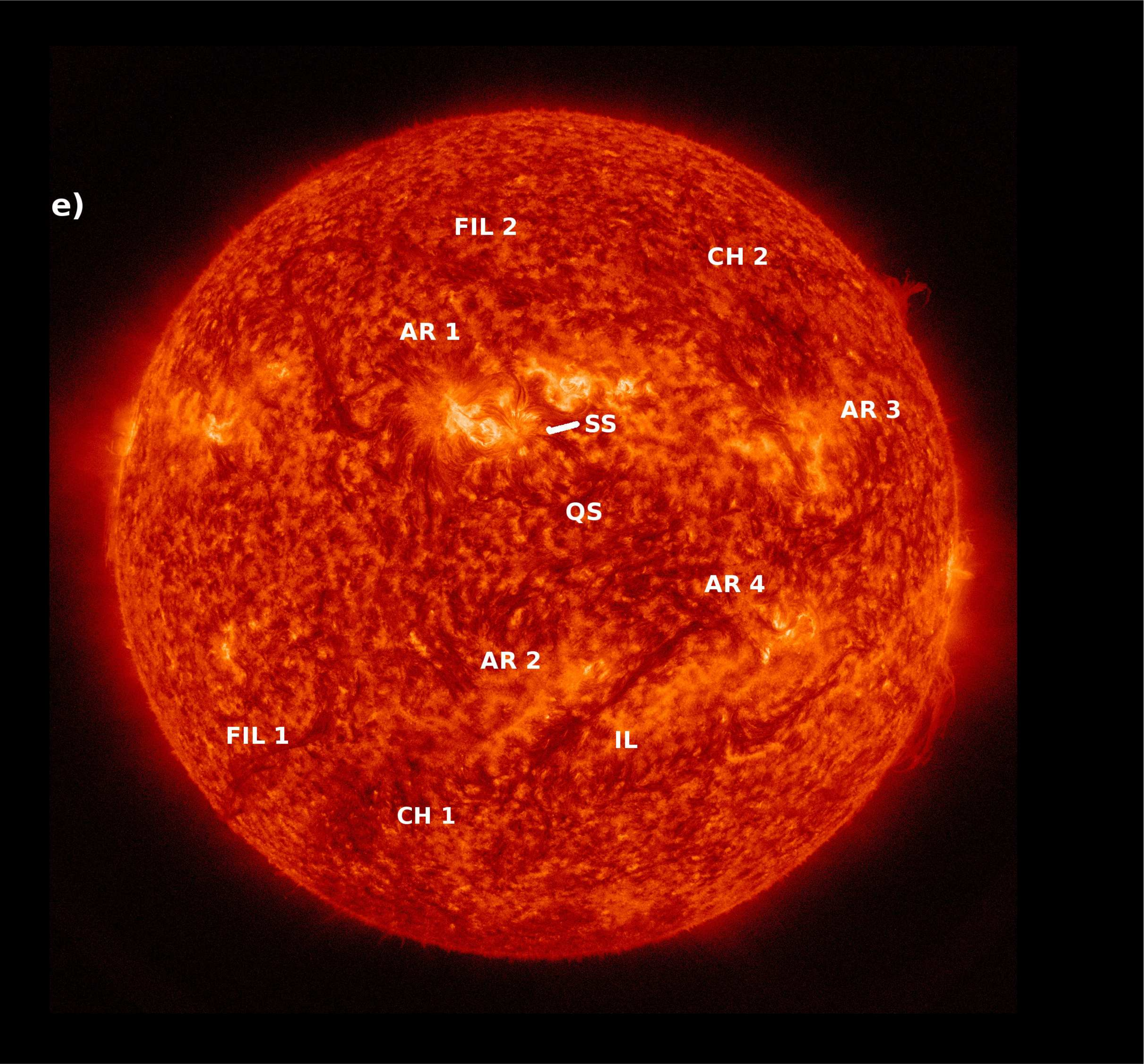}\includegraphics[width=0.45\textwidth]{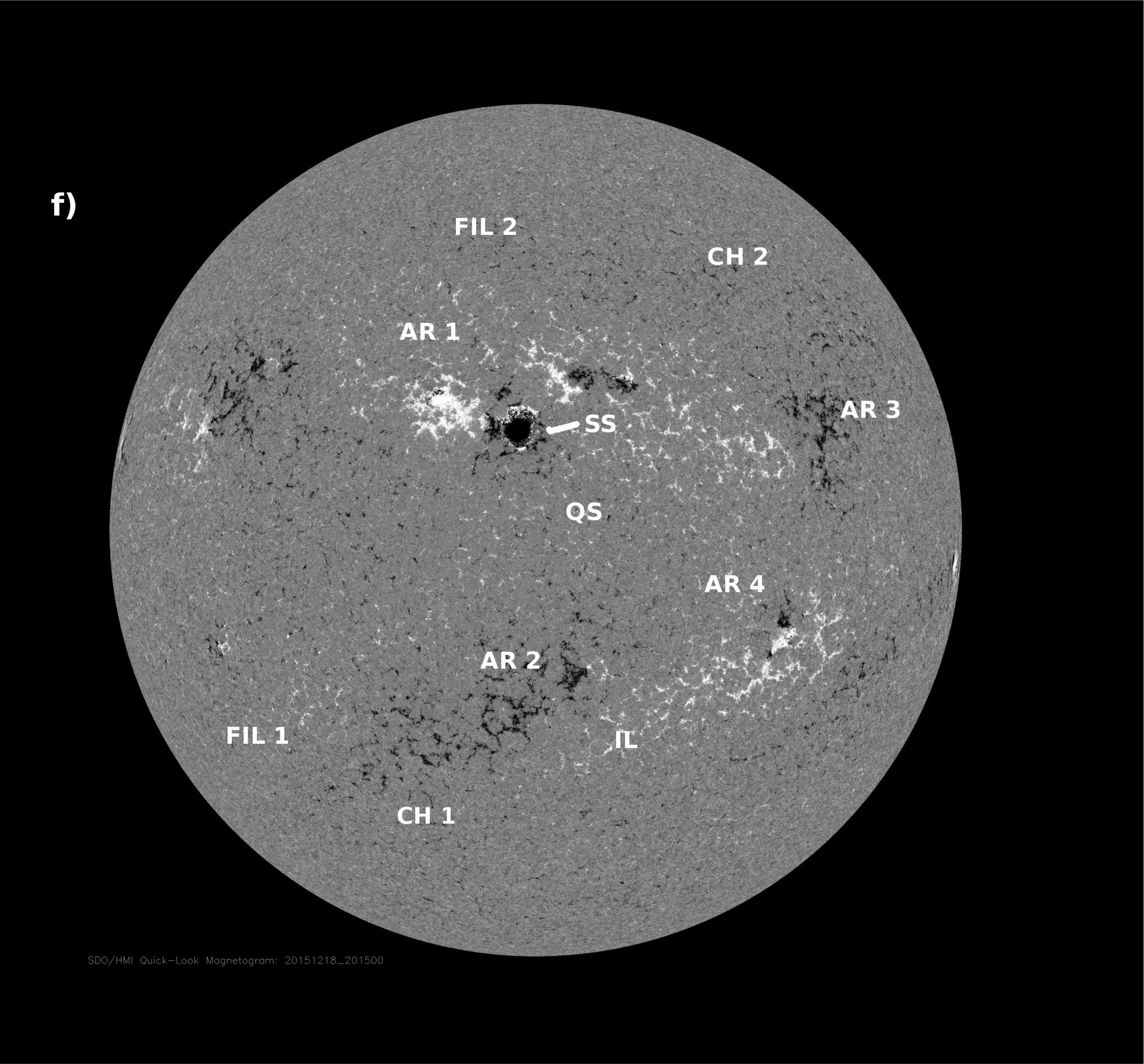}}
\caption{Image of the Sun from different instruments taken on December 18th, 2015. 
{\em Top Left, a)}: ALMA intensity map at 248 GHz ($\lambda = 1.21$ mm, 20h 12m 21s). The brightness temperature 
in K is given on the intensity bar on the right. 
{\em Top Right, b)}: SDO composite image from AIA 30.4 nm, AIA 21.1 nm, AIA 17.1 nm 
instruments (20h 12m 58s UT). 
{\em Middle Left, c)}:  H$\alpha$ filtergram from Cerro Tololo Observatory, NISP (20h 12m UT).
{\em Middle Right, d)}: SDO AIA 170.0 nm filtergram.
{\em Bottom Left, e)}: SDO AIA 30.4 nm filtergram.  
{\em Bottom Right, f)}: SDO HMI magnetic field. 
Several regions of interest are encircled by white lines: 
AR indicates the position of active regions, FIL shows the position of filaments,
SS shows the position of the sunspot, QS shows the position of the central quiet Sun region, IL 
indicates the position of magnetic inversion line, while CH are positions of coronal holes.}
\label{Fig_1} 
\end{figure*}

In the period 2011 -- 2015 Commissioning and Science Verification (CSV) phase of solar 
observations with ALMA took place \citep{Bastian2015, Kobelski2016}. In this frame several solar campaigns were 
performed, lasting typically between one day and one week. Test observational 
material, both single dish and interferometric, was collected and analyzed from which 
solar images were compiled. A first part of the CSV data was recently 
released to the scientific community\footnote{https://almascience.eso.org/alma-data/science-verification}.

An image of the whole solar disc from the December 2015 CSV campaign is used in current analysis. 
In Fig.~\ref{Fig_1}
we present the December 18th, 2015 ALMA image taken with a single dish antenna
together with the Cerro Tololo\footnote{http://halpha.nso.edu}
H$\alpha$ and several SDO\footnote{http://sdo.gsfc.nasa.gov} 
full-disc solar images obtained at the same time. 
Besides the SDO/AIA composite image, the SDO/AIA 170 nm filtergram, the SDO/AIA 30.4 nm 
filtergram and the SDO/HMI magnetogram are presented. 

The full disc solar map was obtained by scanning the solar disc with a 12 m single dish 
total power ALMA antenna (PM03) at a frequency of 248 GHz corresponding to 
$\lambda = 1.21$ mm in a double circle pattern \citep{Phillips2015}. 
The measurement frequency/wavelength belongs to  
Band 6, which is one of the ten ALMA observing bands.
We restricted our analysis to Band 6, because on December 18th, 2015 
no full disc image in the other observing band (Band 3) was available. The observing 
day December 18, 2015 was chosen because of availability of images in other 
wavelength ranges, especially He I 1083 nm, which was not accessible for the other 
observing days.
The measurement started at 20:12:21 UT on December 18, 2015 and lasted for 
about 13 minutes. The beam size amounts to 26 arcsec. 

A two-load method was used to calibrate the data where measurements of 
the known ambient and hot load are used to determine the source 
brightness temperature, together with measurements of zero reference and 
received power on and off the source. 
Also, a correction (multiplication) factor $C$ was 
applied to account for the antenna efficiency: $C = 1.16$ for Band 6 \citep{White2017}.
Data reduction was performed using the Common Astronomy Software Applications (CASA) 
package\footnote{http://casa.nrao.edu}.
The final uncertainty in the brightness temperature is  about 5 \% 
\citep{White2017}.

More information about the measurements, calibration and imaging of the solar observations 
with ALMA for the single dish/total power mode can be found 
in the paper by \citet{White2017}.

In present work we compare the full disc solar image at 1.21 mm obtained on December 18, 2015 
with ALMA  with full disc solar images from the same day in H$\alpha$ line 
(Cerro Tololo Observatory, NISP), in He I 1083 nm line core (NSO SOLIS)\footnote{https://solis.nso.edu/0/index.html}, 
and with various SDO images 
(a composite AIA image at 30.4 nm, 21.1 nm, and 17.1 nm;  AIA filtergrams at 170 nm, at 30.4 nm, and at 19.3 nm; 
 and a HMI magnetogram).

\section{An analysis of the ALMA image and comparison with other data}

A comparison of full-disc solar images taken at different wavelengths 
(Fig.~\ref{Fig_1})
reveals many interesting characteristics. Active regions, which appear bright in 
H$\alpha$ and in EUV, appear also bright at 1.21 mm.  
Filaments on the disc, clearly seen in H$\alpha$ can not easily 
be identified at 1.21 mm; their brightness temperature is obviously rather 
close to the quiet Sun background.  Finally, coronal holes which are identified as 
dark regions in EUV have no visible counterparts at 1.21 mm (no significant change of 
the brightness 
temperature compared to the quiet Sun background). 

In the solar ALMA image (Fig.~\ref{Fig_1}a) several regions of interest are indicated.
The brightness 
temperature levels, measured for these regions of interest, are presented in Table~\ref{table_1}. 
Brightness temperatures of various structures are calculated as the mean values of all pixels within 
regions of interest denoted in Fig.~\ref{Fig_1}a. 
The quiet Sun levels are determined from circles centered at various distances, $r$, from the solar 
disc center. At the solar disc center radius of the circle is 5 pixels, while for all other quiet Sun measurements 
it is 30 pixels. The size of a pixel is about 3 arc sec. The coronal hole level is 
measured within the circle of a radius of 30 pixels located within the CH1. 
The contours of active region are defined using a numerical criterion: all pixels having 
brightness temperature $T_b > 6900$ K in the area of the AR1 are included in the region of 
interest. With this criterion the bright area associated with active region is included in the 
region of interest and sunspots are automatically excluded from the active region area. 
The sunspot is also defined with the numerical criterion: all pixels with $T_b < 6400$ K at the 
position denoted with SS. 
We note that this selection criteria (for AR1 and SS) are crude and to some extent arbitrary. 
The level of magnetic inversion line is defined as the mean value 
within the rectangle within the inversion line area (IL), while the prominence level was 
taken from the shape of the H$\alpha$ image put over. 
The number of pixels used in the averaging process 
depends on the size of a structure and varies 
from $n = 78$ in the central quiet Sun region to $n = 3048$ in the 
active region.

To quantitatively measure the brightness temperature of the above mentioned structures the 
center-to-limb function should be taken into account. As we can see from Table~\ref{table_1}, there is 
a clear limb brightening at 1.21 mm (compare the brightness temperatures of the quiet Sun at different 
radial distances from the solar disc center). 
The brightness temperature values of various solar structures are compared with the quiet Sun 
levels measured at the same radial distances from the solar disc center as the respective structures 
but along the solar equator in the east direction 
(Table~\ref{table_1}). Finally, the relative brightness temperatures are given in Table~\ref{table_1}: a positive 
(negative) sign indicates that the feature is brighter (darker) than the surroundings.

\begin{table*}
\centering
\caption{Brightness temperatures $T_b$ of various structures in the solar atmosphere. } 
\label{table_1} 
\begin{tabular}{c c c c c c c}
\hline\hline
Structure & $r$         & $T_b {\rm (QS)}$ & $n$(QS) & $T_b {\rm (structure)}$  & $n$(structure) & $\Delta T_b=T_b {\rm (structure)}-T_b {\rm (QS)}$  \\
                & (pixels)  &    (K)                     &  (pixels)  &  (K)                                &  (pixels)           &    (K)                \\
\hline
QS             &   0         &  $6040 \pm  70$   &        78    &   $6040 \pm 70$          &         78              &       $0$           \\ 
SS             &   77       &  $6170 \pm 140$   &     2833   &   $6080 \pm 210$       &       136              &      $-90$         \\ 
AR1           &   105     &  $6240 \pm 150$   &     2832   &   $7250 \pm 210$       &      3048             &   $+1010$       \\ 
IL               &   147     &  $6300 \pm 160$   &     2833   &   $6130 \pm 160$       &         897            &    $-170$         \\
PR (FIL1)   &   229    &  $6460 \pm 160$   &     2824   &   $6350 \pm 110$        &        154             &    $-110$          \\
CH1          &  273     & $6590  \pm 140$   &     2833    &  $6540 \pm 130$        &        2804           &    $-50$           \\
\hline
\end{tabular}
\tablefoot{
Structures refer to the regions of interest denoted in Fig.~\ref{Fig_1}a: 
central quiet Sun (QS), 
sunspot (SS), active region (AR1), magnetic inversion  line (IL), prominence on the disc (PR/FIL1) and 
coronal hole (CH1). The mean values of brightness temperature 
of these structures, given here together with their corresponding standard deviations, 
are compared with the ones of the quiet Sun regions at the same radial distances from 
the center of the solar disc, $r$. The averaging was performed within selected regions of interest 
and the number of pixels within is denoted by $n$. 
}
\end{table*}

\begin{figure*}
\centerline{\includegraphics[width=0.45\textwidth]{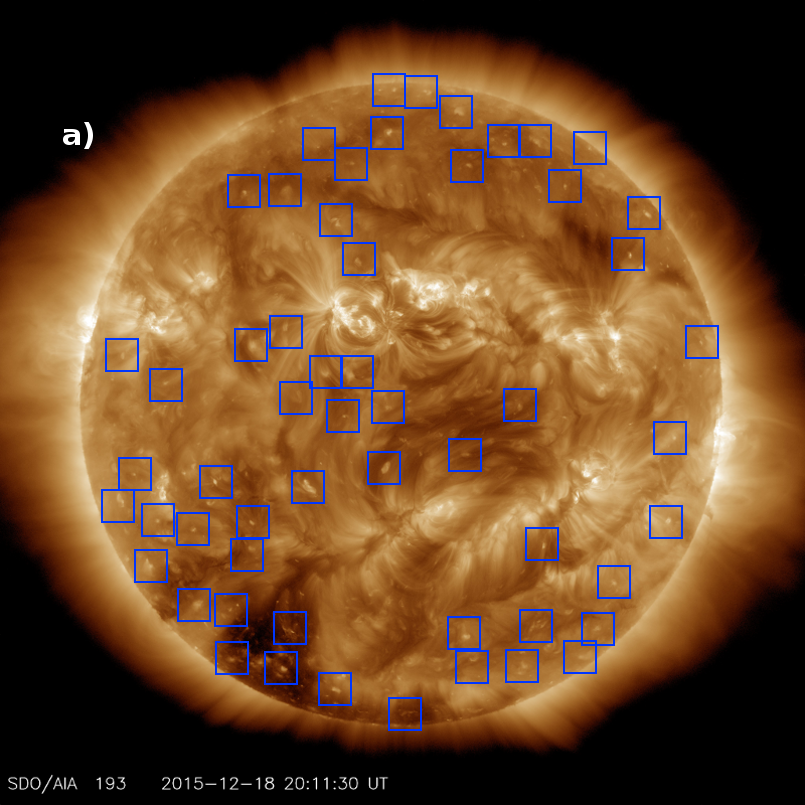}
\includegraphics[width=0.45\textwidth]{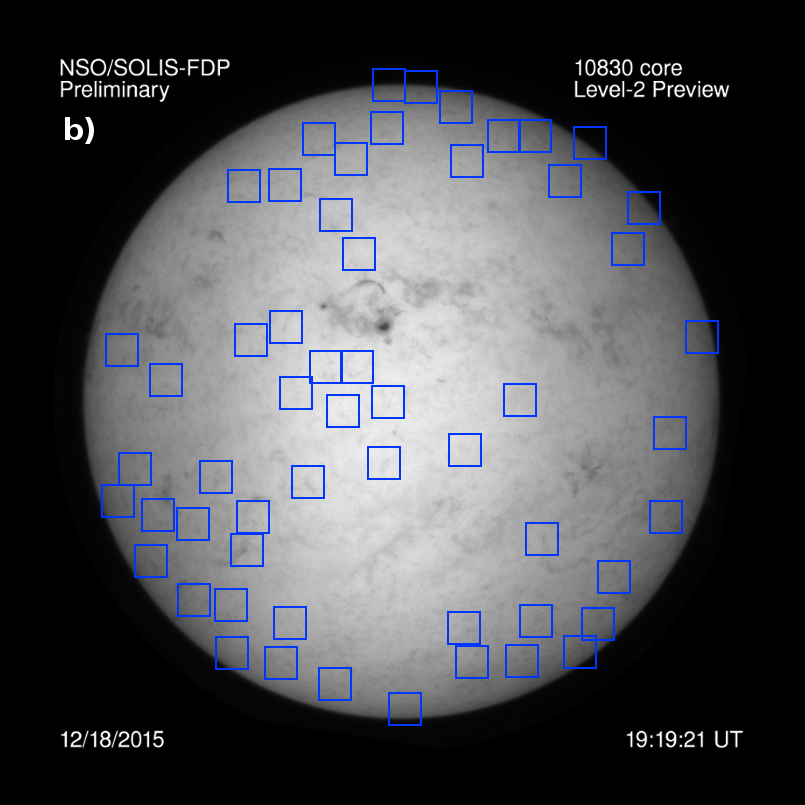}}
\centerline{\includegraphics[width=0.75\textwidth]{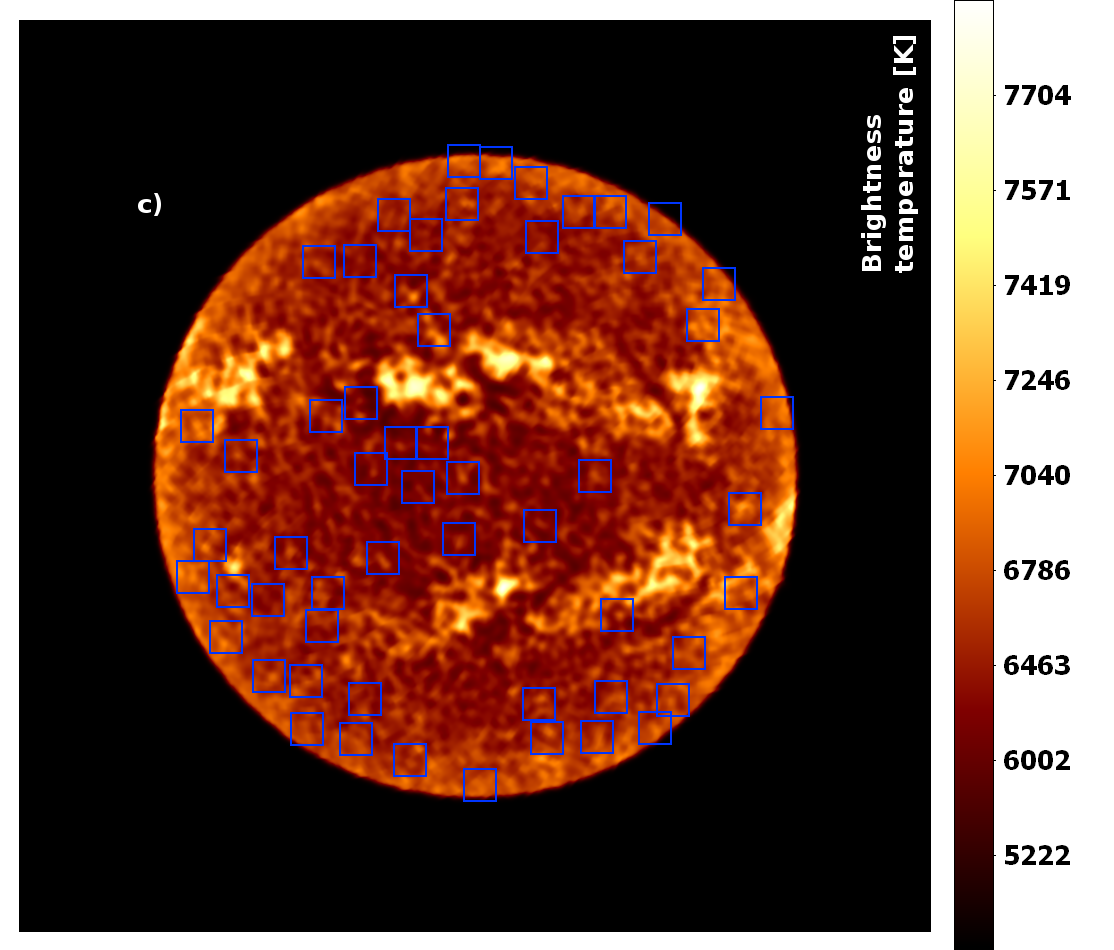}}
\caption{Image of the Sun taken at 3 different wavelengths on December 18th, 2015 at the same time as Fig.~\ref{Fig_1}.
Coronal bright points are denoted with blue boxes which are at the 
same positions in all three images.  
{\em Top Left, a)}: SDO AIA 19.3 nm filtergram.  
{\em Top Right, b)}: NSO SOLIS He I 1083 nm core filtergram.    
{\em Bottom, c)}:  ALMA intensity map at 248 GHz ($\lambda = 1.21$ mm). 
}
\label{Fig_2} 
\end{figure*}

\begin{figure*}
\centerline{\includegraphics[width=0.55\textwidth]{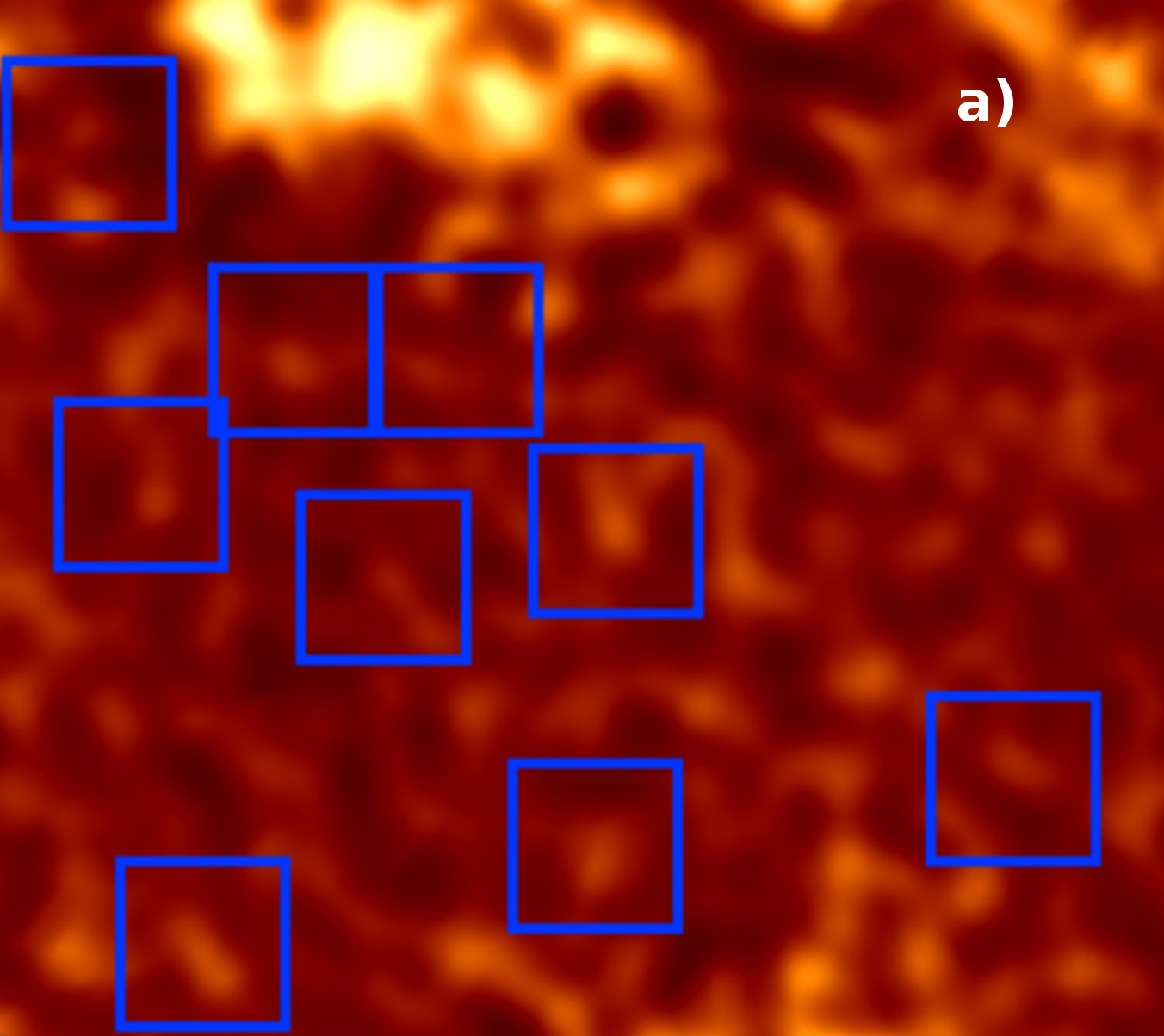}}
\centerline{\includegraphics[width=0.55\textwidth]{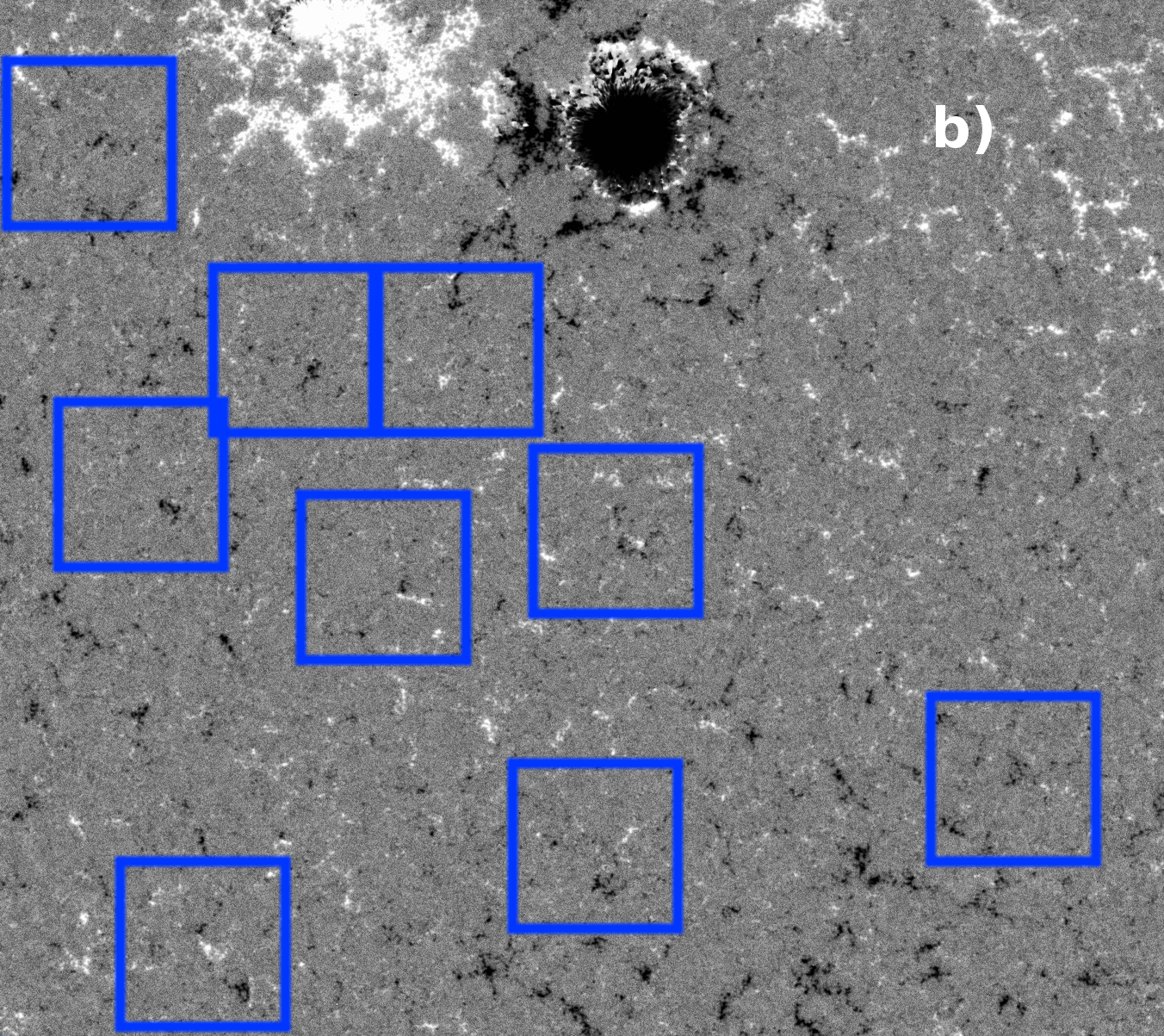}}
\caption{A zoomed in portion of Fig.~\ref{Fig_2}, between the large sunspot and solar equator, here presented 
magnified. The horizontal length is about 490 Mm. 
{\em Top, a)}:   ALMA intensity map at 248 GHz ($\lambda = 1.21$ mm).
{\em Bottom, b)}:    SDO HMI magnetic field.
}
\label{Fig_3} 
\end{figure*}

\subsection{Quiet Sun}

It is not easy to define and observationally identify the quiet Sun regions at mm wavelengths 
because of 
highly complex fine structure chromosphere seen (Fig.~\ref{Fig_1}a). However, we have 
chosen 6 regions of interest representing the quiet Sun areas at different radial distances 
from the solar disc center, see Table~\ref{table_1}.  
The quiet Sun brightness temperature increases steadily with increasing radial distance
from the disc center consistent with a pronounced limb brightening. 
The amount of this limb brightening is in the order of 10 \% in agreement with recent results of 
\citet{Alissandrakis2017}. 
The central brightness temperature of the quiet Sun is measured as the average value 
of all intensities within the small circle located at the center of the solar disc in Fig.~\ref{Fig_1}a. 
This value is 6040 K (Table~\ref{table_1}) fully consistent with the central quiet Sun 
brightness temperature measured at 1.21 mm for the December 2015 data  
\citep{White2017}.

\subsection{Active regions and sunspots}

Active regions are clearly bright at the wavelength of 1.21 mm (Fig.~\ref{Fig_1}a). 
They are also bright in other images of the solar chromosphere and corona: SDO AIA 
composite image (Fig.~\ref{Fig_1}b), H$\alpha$ (Fig.~\ref{Fig_1}c), SDO AIA 170 nm 
(Fig.~\ref{Fig_1}d), and SDO AIA He I 30.4 nm (Fig.~\ref{Fig_1}e). As it is well known 
they lie over strong magnetic field regions in the solar photosphere (Fig.~\ref{Fig_1}f). 
We now analyze in more detail the most prominent active region, denoted as AR1 in Fig.~\ref{Fig_1}.
This active region (AR1) has an average magnetic field strength of about 1000 Gauss (as measured in the 
magnetogram, Fig.~\ref{Fig_1}f) and  
the average measured brightness temperature (Fig.~\ref{Fig_1}a) is given in 
Table~\ref{table_1}. In defining the contours of the AR1 the sunspots are 
excluded, because the sunspot umbras have a lower brightness temperature then the 
rest of the AR.  
When compared with the average brightness temperature of a 
quiet Sun region at the corresponding radial distance from the disc center (Table~\ref{table_1}), 
the AR1 has an average surplus emission of more than 1000 K.  

Although active regions are bright at 1.21 mm, sunspots are clearly dark, especially 
when compared with active region environment (Fig.~\ref{Fig_1}). 
We measured the average brightness temperature of the large sunspot in AR1 (Fig.~\ref{Fig_1}a)
and its intensity (6080 K) is lower not only than the intensity of  the surrounding active region, but also 
than the average brightness temperature of the quiet Sun at the corresponding radial distance 
from the solar disc center  (Table~\ref{table_1}). Finally we note that the large sunspot SS (Fig.~\ref{Fig_1}a) 
clearly corresponds to the area of a strong magnetic field (Fig.~\ref{Fig_1}f) in the order of 1500 - 2500 Gauss.

\subsection{Prominences on the solar disc}

On the ALMA observing day, December 18, 2015, several prominences on the solar disc were observed. 
They can be identified as dark filaments seen in the H$\alpha$ filtergram presented in 
Fig.~\ref{Fig_1}c. We now search for their counterparts in the ALMA map (Fig.~\ref{Fig_1}a) and 
find that no distinctive structures are present at the places of H$\alpha$ filaments in the 1.21 mm map. 
To quantitatively analyze the radiation intensity of filaments on the disc the brightness temperature of a part 
of the ALMA map (Fig.~\ref{Fig_1}a) at the position of the filament FIL1  (Fig.~\ref{Fig_1}c) was measured. 
The result is presented in Table~\ref{table_1} together with the brightness temperature of a quiet Sun regions 
measured at the same radial distance from the solar disc center as the FIL1. At the wavelength of 1.21 mm
the prominence on the disc has a brightness temperature lower than the quiet Sun level by slightly more 
than 100 K. 
The reliability of this result depends on the overlying precision (in the order of 5 arc sec) and the 
deviations from the measured mean intensity value (Table~\ref{table_1}).

\subsection{Magnetic inversion lines}

We now turn to the following question:  In the 1.21 mm map, what are the dark, large, elongated structures lying    
between active regions (Fig.~\ref{Fig_1}a)? By comparison with the SDO HMI magnetogram (Fig.~\ref{Fig_1}f), it can 
be seen that all of these ALMA structures are cospatial with the inversion lines of the large scale magnetic 
field on the Sun. For a quantitative analysis a part of the magnetic inversion line was chosen (Fig.~\ref{Fig_1}a) 
and its brightness temperature was measured and compared with the corresponding quiet Sun region 
(Table~\ref{table_1}). The net deficit of brightness temperature of the measured portion of the magnetic 
inversion line is  $-$  170 K.  Finally, it is also noteworthy that all these large dark ALMA structures are cospatial 
with dark structures visible in He 30.4 filtergram (Fig.~\ref{Fig_1}e).

\subsection{Coronal holes}

Coronal holes can be most easily identified in the EUV images of the Sun (Fig.~\ref{Fig_1}b and Fig.~\ref{Fig_2}a). 
On the observing day, December 15, 2015, two large coronal holes were present on the Sun, denoted by CH1 and CH2. 
By comparing the EUV images with the ALMA map (Fig.~\ref{Fig_1}a) it is easily seen that coronal holes have no 
distinctive counterparts at 1.21 mm. Indeed, the average brightness temperature measured for a region of interest 
within the coronal hole CH1 (Fig.~\ref{Fig_1}a) is only 50 K below the adjacent quiet Sun region identified at the 
same radial distance from the solar disc center (Table~\ref{table_1}).
The CH1 is the most prominent solar coronal hole on the observing day, so only it was measured 
as a representative example.

\subsection{Coronal bright points}

Coronal bright points are small bright structures belonging to the low corona which can be clearly identified 
in X-ray and EUV images of the Sun. These structures have dimensions in the order of 15 - 30 arc sec and 
are closely related to the underlying small-scale magnetic elements and He I 1083 nm dark points 
\citep{Harvey-Angle1993, Brajsa2002}.
Their visual classification put them into one of the three classes: point-like structures, small loops and small 
active regions \citep{Brajsa2002}. Several dozens of identified coronal bright points can be seen in Fig. 1 in the 
paper by \citet{Sudar2015}.  Coronal bright points are most easily observed outside active regions.  

A solar filtergram recorded with SDO/AIA at 19.3 nm is presented in Fig.~\ref{Fig_2}a and 56 coronal bright 
points (only point-like structures and small loops) are identified and denoted with blue boxes. 
 In Fig.~\ref{Fig_2}b and Fig.~\ref{Fig_2}c the He I 1083 nm 
filtergram and the ALMA 1.21 mm map are presented, respectively, with the above mentioned 56 blue boxes 
placed in the same positions as in the SDO/AIA image in Fig.~\ref{Fig_2}a. 
By a careful and detailed comparison of the three images we can conclude: 
(i) the great majority (75 \%) of all coronal bright points from the 
EUV image corresponds to He I 1083 nm dark points, (ii) the great majority (82 \%) of all coronal bright points from 
the EUV image corresponds to the ALMA 1.21 mm bright points, 
and (iii) the shape and orientation of the structures is preserved and reproduced 
in all three spectral ranges. These inferences have two important implications: (a) coronal bright points have almost 
a one to one 
correspondence with the ALMA 1.21 mm bright points and (b) this is an important and conclusive proof that 
the precision of ALMA can be best tested with coronal bright points and this spatial precision is 
in the order of 5 arc sec.

Finally, to investigate in more detail coronal bright points, which we can now also call ALMA bright points, 
we present a zoomed in portion of the ALMA image together with a corresponding magnetogram in Fig.~\ref{Fig_3}. 
By comparing these two images following important conclusion can be made: (i) ALMA bright points are clearly 
related to small-scale bipolar magnetic regions (also one polarity is prevailing in some cases) and 
(ii) following the shape 
of magnetic fine structure the pattern is preserved and reproduced in the ALMA image: small bright structures 
identified in the ALMA image (Fig.~\ref{Fig_3}a) are co-spatial with small-scale magnetic features (Fig.~\ref{Fig_3}b) and 
the mm brightness is clearly proportional to the magnetic field strength (for places on solar surface outside 
sunspots). 

\section{Summary, discussion and conclusions}

We now summarize observational results for the solar objects analyzed in previous Sect.,  compare them with other 
existing studies and list the most important conclusions of the present study.  

\bf Quiet Sun. \rm 
The definition and identification of the quiet Sun regions and the 
determination of the brightness temperature of the quiet Sun level  is not easy, which 
is primarily due to the highly complex fine structure of the solar chromosphere 
\citep{Wedemeyer2016} and a frequency dependent 
center to limb intensity function with a pronounced limb brightening at mm wavelengths \citep{Bastian1993a, Brajsa1994, 
White2017, Alissandrakis2017}. 
In present analysis we have determined the brightness temperature of the quiet Sun region at 
the disc center to be 6040 K at 1.21 mm, which is fully consistent with the central brightness temperature 
determined  by \citet{White2017}  for  the 2015 data  (6040 $\pm$ 250 K). 
Also the estimated limb brightening at 1.21 mm is up to 10 \%, consistent with the results of 
\citet{Alissandrakis2017}.

\bf Active regions and sunspots. \rm At the wavelength of 1.21 mm active regions are clearly bright, as expected and 
in accordance 
with all previous studies for the mm wavelength range. The strong excess emission measured in this work 
amounts to more than 1000 K for  the analyzed active region. However, sunspots are seen in absorption, as 
localized dark areas, in agreement with some earlier studies for the wavelengths $\lambda \le 3.5$ mm 
\citep{Lindsey1995, Loukitcheva2014, Iwai2015}. 
As present analysis shows, the analyzed sunspot has brightness temperature lower by 90 K than a 
quiet Sun region at the same radial distance from the solar disc center allowing us to conclude that 
sunspots are darker not only in comparison with ambient active regions, but also to the quiet Sun regions.

\bf Prominences on the solar disc. \rm As we have seen, prominences on the solar disc can not be 
easily identified against the surrounding quiet Sun background. However, the quantitative analysis 
reveals that prominences have brightness temperatures lower in the order of 100 K than the quiet 
Sun regions at the same radial distances from the disc center at the observing wavelength of 1.21 mm. 
This is in agreement with a previous study \citep{Bastian1993b} in which a negligible contrast, 
but still a lower brightness temperature than the quiet Sun level,  was measured 
for prominences on solar disc at the wavelength near 1 mm. 
The appearance (emission vs. absorption) of prominences on the solar disc at mm and sub-mm wavelengths 
is rather frequency dependent and further multifrequency ALMA observations are needed to 
bring more insight into this interesting topic.

\bf Magnetic inversion lines. \rm  In the ALMA map taken at 1.21 mm the inversion lines of the large scale 
magnetic field on the Sun are clearly visible as large elongated dark structures. The average brightness 
temperature of the measured part of the inversion line has a brightness temperature  170 K below the 
quiet Sun level. This is in agreement with earlier studies performed at mm wavelengths. 
\citet{Brajsa1992} and \citet{Vrsnak1992} performed an extensive statistical analysis of full disc solar 
maps at the wavelength of 8 mm and concluded that 99$\%$ of all dark structures lie over the inversion 
lines of the large scale magnetic field. \citet{Bastian1993b} analyzed full disc solar maps at the 
wavelength near 1 mm and also found a lower brightness temperature at the places of magnetic 
inversion lines. A clear association between large scale  dark structures and magnetic inversion lines 
can also be found in the recently published solar ALMA maps \citep{White2017}.
This observationally found correlation between regions of lower brightness temperature 
at mm wavelengths and magnetic inversion lines can be interpreted in terms of reduced heating and a 
smaller density scale 
height in the chromosphere along the inversion lines of the large scale magnetic field where the magnetic field 
vector is horizontal \citep{Vrsnak1992, Bastian1993b, Brajsa1993}.

\bf Coronal holes. \rm Similarly as prominences on the disc, coronal holes have no 
discernible contrasts against the quiet Sun background at the observing wavelength of 
1.21 mm. The net intensity depression measured in present work is only  50 K.
This is in general agreement with previous studies mentioned in Sect. 1, but there is 
an obvious lack of other coronal holes observations at the wavelength around 1 mm. 
So, further multifrequency observations with ALMA are desirable, taking into account 
that coronal contribution, which is dominantly responsible for the radiation of coronal 
holes, is becoming less and less important for  shorter wavelengths.

\bf Coronal bright points. \rm In this work we presented clear observational evidence 
that the great majority of all coronal bright points from the EUV image corresponds to the He I 1083 nm dark points 
(75 \%) 
and to the ALMA 1.21 mm bright points (82 \%). 
Moreover, all ALMA 1.21 mm bright points show well-defined 
relationship with magnetic structures (100 \% correspondence),  
mostly with small scale bipolar magnetic regions and in some 
cases with small unipolar magnetic regions. 
A correlation in position of up to 50$\%$ between 20 cm bright points and coronal bright points  was found 
by \citet{Habbal1988} and by \citet{Nitta1992}. A similar result for the wavelength of 17.6 mm was reported 
by \citet{Kundu1994}.These results are very interesting, but we note that the observing 
wavelength of 20 cm is two orders of magnitude and the observing wavelength of 17.6 mm is 
one order of magnitude  larger than the one used here (1.21 mm). 
It is also important to note that the 20 cm emission comes from the solar corona and 
the 1.21 mm radiation originates in the solar chromosphere: these are two different layers in the solar 
atmosphere having rather different temperatures. 
An attempt to identify a coronal bright point in a 8 mm solar map failed due to the poor spatial 
resolution \citep{Brajsa2007b}. Coronal bright points are magnetic structures associated with 
bipolar magnetic elements: ephemeral (emerging) regions or cancelling magnetic features 
\citep{Brajsa2004, Sudar2015}. Best identification of coronal bright 
points is in EUV images of the Sun (e.g. with the SDO) and, when the spatial resolution permits, it can be 
seen that they consist of small loops in almost all cases.  The observed higher brightness temperatures 
of coronal bright points in the ALMA map at 1.21 mm is due to the locally increased temperature and 
density in these loops. 

In present work we analyzed a full disc solar ALMA map and performed a detailed comparison 
of this map with solar images obtained in other relevant wavelength ranges. The detectability of  
various solar features was checked in both directions (an identification of solar structures known at 
other wavelengths in the ALMA image and an identification of prominent ALMA structures in images 
at other wavelengths). 
It was found that the ALMA image was oriented correctly and that the procedure of overlay and co-align  the 
ALMA image with other images was performed properly. 
In subsequent papers we compare the observational 
results from this paper with the theoretical models (Paper II) and later investigate chromospheric fine structure 
in more detail using ALMA interferometric measurements.

\begin{acknowledgements}Research leading to this work was performed within ESO 
Development Plan Study: Solar Research with ALMA (2014 - 2017). 
This work has been supported in part by Croatian Science Foundation under the project 6212 
''Solar and Stellar Variability'' and by the European Commission FP7 project SOLARNET (312495, 2013 - 2017), 
which is an Integrated Infrastructure Initiative (I3) supported by FP7 Capacities Programme.
This paper makes use of the following ALMA data: ADS/JAO.ALMA\#2011.0.00020.SV.
ALMA is a partnership of ESO (representing its member states), NSF (USA) and
NINS (Japan), together with NRC (Canada) and NSC and ASIAA (Taiwan), and KASI
(Republic of Korea), in cooperation with the Republic of Chile. The Joint ALMA
Observatory is operated by ESO, AUI/NRAO and NAOJ.
We are grateful to the ALMA project for making solar observing with ALMA possible.
SDO is the first mission launched for NASA\'\,s Living With a Star (LWS) Program.
This work utilizes GONG data obtained by the NSO Integrated Synoptic Program (NISP), 
managed by the National Solar Observatory, the Association of Universities for Research in 
Astronomy (AURA), Inc. under a cooperative agreement with the National Science Foundation. 
The data were acquired by instrument operated by the Cerro Tololo Interamerican Observatory.
The National Radio Astronomy Observatory is a facility of the National 
Science Foundation operated under cooperative agreement by Associated 
Universities, Inc.
The authors would like to thank T. Bastian, the PI of the North American ALMA Solar Development Plan under which much of 
the testing and commissioning work was coordinated.
SW acknowledges support by the SolarALMA project. This project has received funding from the European Research Council (ERC) 
under the European Union\'\,s Horizon 2020 research and innovation programme (grant agreement No. 682462).
The trip of YY to 2015 ALMA Solar Campaign was partially supported by NSFC grant 11433006. 
MS was supported by JSPS KAKENHI Grant Number JP17K05397.
Finally, the authors would like to thank R. Hills, K. Iwai, and M. Loukitcheva for helpful comments and suggestions. 
\end{acknowledgements}

\bibliographystyle{aa}
\bibliography{alma_csv_obs_v1} 

\end{document}